\documentclass[12pt,preprint]{aastex}

\lefthead{Itoh et al.}
\righthead{}

\begin{document}

\title{Coronagraphic Search for Extra-Solar Planets around 
$\epsilon$ Eri and Vega
\footnote{Based on data collected at the Subaru Telescope, 
which is operated by the National Astronomical Observatory of Japan.}}
\author{Yoichi Itoh\altaffilmark{2},
Yumiko Oasa\altaffilmark{2},
and
Misato Fukagawa\altaffilmark{3},
}
\altaffiltext{2}{Graduate School of Science and Technology, Kobe University,
1-1 Rokkodai, Nada, Kobe, Hyogo 657-8501, Japan, 
yitoh@kobe-u.ac.jp}
\altaffiltext{3}{Graduate School of Science, Nagoya University,
Furo-cho, Chikusa-ku, Nagoya 464-8601, Japan}

\begin{abstract}
We present the results of a coronagraphic imaging
search for extra-solar planets around the young main-sequence stars,
$\epsilon$ Eri and Vega.
Concentrating the stellar light into the core of the point
spread function by the adaptive optic system
and blocking the core by the occulting mask in the coronagraph,
we have achieved the highest sensitivity for point sources in close vicinity of
the both central stars.
Nonetheless we had no confidential detection of a point source around the stars.
The observations give the upper limits on the masses of the planets to
4 $\sim$ 6 Jupiter mass and 5 $\sim$ 10 Jupiter mass
at a few arcsecond from $\epsilon$ Eri and Vega, respectively.
Diffuse structures are also not detected around both stars.
\end{abstract}

\keywords{
stars: individual ($\epsilon$ Eri, Vega) --- 
techniques: high angular resolution ---
planetary systems.
}

\section{INTRODUCTION}

Searches for extra-solar planets are very successful;
over 180 planets have been discovered.
However all discoveries so far were made using indirect methods, i.e. Doppler 
shift measurements and the transit method.
Direct detection --- imaging of the emission from an extra-solar
planet --- will open the door to investigating 
chemistry, meteorology, and biology
under conditions completely different from those of the Earth and
the other planets in the Solar System.

The small angular separation between a planet and a central star
and the huge difference in the brightness of the two objects make
direct imaging difficult.
The direct detection of extra-solar planets around
pre-main sequence stars is less challenging in terms
of the brightness difference,
since young planets are bright in radiation.
Itoh et al. (2005) discovered a young brown dwarf companion to the classical 
T Tauri star DH Tau, 
and the companionship was established by proper motion measurements.
They derived the effective temperature by comparing its near-infrared spectrum
with synthetic spectra of young low-mass objects (Tsuji et al. 2004).
The mass is estimated to be 30 -- 50 Jupiter mass ($M_{\rm J}$)
by comparison to evolutionary tracks (Baraffe et al. 2003; D'Antona et al. 1997)
on the HR diagram.
Chauvin et al. (2004) presented a direct image of a giant-planetary mass 
object around a young brown dwarf in the TW Hya association.
Neuh\"auser et al. (2005) announced the discovery of
a proto-planet around GQ Lup.
They estimated its mass to be 1 -- 40 $M_{\rm J}$ using evolutionary
tracks of Wuchterl et al. (2005) and Baraffe et al. (2003).
However, in general, mass estimates of a proto-planet have large uncertainty. 
First, as pointed out by Itoh et al. (2005),
determining the effective temperature by comparison of the spectrum
to the spectra of field dwarfs tends to underestimate the value.
Moreover, the evolutionary tracks of low-mass objects have large uncertainties.
For example, we derive the mass of DH Tau B to be only 5 $M_{\rm J}$,
i.e. in the planetary mass range, if using the evolutionary track of Wuchterl et al.
(2005).
These two factors make it still
unclear whether the low-luminosity companions are proto-planets.

Another approach to direct detection of extra-solar planets is, of course, 
detection of an extra-solar planet around a main-sequence star.
Such an object does not have large ambiguities in the effective temperature 
estimate and in the mass estimate on the HR diagram.
However, such an object is no longer bright in radiation.
The flux ratio of the reflection light of a planet to a central star is
described as
$\frac{F_{\rm p}}{F_{\rm s}} = \frac{A}{2} (\frac{R_{\rm p}}{2a})^{2}$,
where $A$, $R_{\rm p}$, and $a$ are albedo, the radius, and 
the semi-major axis of the planet, respectively.
Reflection lights from all the extra-solar planets discovered by the 
Doppler shift measurements or the transit methods are difficult 
to be directly detected in reflected light,
due to their faintness ($>$ 21 mag at the $H$-band) and/or due to their
small separation ($<$ 0.3\arcsec) from the star.

Alternative targets for direct detection are unknown extra-solar planets
around nearby young main-sequence stars.
Such planets are expected to be still bright in a contraction phase.
Here we report the results of the coronagraphic observations
of extra-solar planets around such young dwarfs, $\epsilon$ Eri and Vega.
These stars are surrounded by dust rings, suggestive not only of
their youths but also of the presence of a planet between the star and 
the ring.

\section{TARGETS}

\subsection{$\epsilon$ Eri}

$\epsilon$ Eri is the nearest young dwarf ($d\sim 3.3$ pc)
with the indication of an extra-solar planet
given by Doppler shift measurements
(Hatzes et al. 2000). 
Amplitude of the Doppler shift is consistent with a planetary
mass companion with $a = 3.4$ AU (1\farcs0).
With its high chromospheric activities, $\epsilon$ Eri is believed to be 
young ($\sim$ 730 Myr; Song et al. 2000).

A clumpy debris disk is discovered around $\epsilon$
Eri in the submillimeter wavelengths (Greaves et al. 1998).
The disk has a ringlike morphology with a peak at 60 AU ($18\arcsec$) 
from the central star and a cavity within 30 AU ($9\arcsec$). 
The inclination of the disk ($i$) is estimated to be $\sim$ 25\arcdeg.
If the planet suggested by the Doppler shift measurement orbits
coplanar with the debris disk, 
its mass is $\sim 2 M_{\rm J}$.
The $H$-band apparent magnitude of a 2 $M_{\rm J}$ planet 
is estimated to be 22.2 at 500 Myr and 25.4 at 1 Gyr (Baraffe et al. 2003).

The structure of the disk is also suggestive of the existence of
a giant planet in orbit with
moderate semimajor axis ($40\sim60$ AU; Ozernoy et al. 2000; 
Quillen \& Thorndike 2002).
Kokubo \& Ida (2002) predict that multiple giant planets may form
in a moderate-mass disk if the disk has
a long dissipation timescale. 
Both facts, that the
circumstellar disk is long-lived and that a
giant planet may orbit at 3.4 AU,
imply the existence of other giant planets in the outer region.

The factors above combine to make $\epsilon$ Eri
an attractive target for direct detection
of an extra-solar planet.

\subsection{Vega}

The distance and age of this star are 7.76 pc and 350 Myr, respectively
(Song et al. 2001).
An extended circumstellar disk is found by sub-millimeter observations
(Holland et al. 1998).
It has two dust emission peaks
at 60 AU (8\farcs0) and 75 AU (9\farcs5)
from the central star.
(Koerner et al. 2001; Wilner et al. 2002).
From the dust distribution,
Ozernoy et al. (2000) predicted a 2 $M_{\rm J}$ planet with 
the semi-major axis of $50\sim60$ AU.
At the distance of Vega,
the apparent $H$-band magnitude of a $2 M_{\rm J}$ planet is
estimated to be 19.2 and 24.1 for age of 100 Myr and 500 Myr, respectively
(Baraffe et al. 2003). 

Because of the pole-on geometry suggested by the circular symmetric structure
of the dust and because of the spectral type of A0, 
a planet is difficult to be detected by Doppler shift measurements,
if any.

\section{OBSERVATIONS AND DATA REDUCTION}

Because an extra-solar planet is thought to be
faint and located in close vicinity of a bright central
star, observations with high sensitivity and high dynamic range are strongly
required. 
A stellar coronagraph with an adaptive optics system is one of
the instruments suitable for such observations.

Coronagraphic observations of $\epsilon$ Eri and Vega
were carried out on 2003
November 08 and 09 under fair condition with occasional cirrus.
We used CIAO,
which is equipped with a 1024$\times$1024 InSb Alladin~II detector
with a spatial scale of 0\farcs0213 pixel$^{-1}$.
The observational band is the $H$-band. 
Because of the low effective temperature of a planet,
we expect the thermal emission in the $K$-band to be
suppressed below detection limit by atmospheric methane absorption.
For example, the apparent $K$-band magnitude of a $5 M_{\rm J}$ planet
around $\epsilon$ Eri is estimated to be 22 $\sim$ 26 mag, 
several magnitude fainter than its $H$-band magnitude.
The spatial
resolution provided by the adaptive optics system was 0\farcs07 (FWHM)
for the natural seeing of $\sim$0\farcs5.  
The occulting masks were made of
chrome on a sapphire substrate, within
which the transmittance was a few tenths of a percent.  
This allowed
us to measure the accurate position of the central object.  
Occulting masks with diameters of 1\farcs0 and 2\farcs0
were used for $\epsilon$ Eri and Vega, respectively.
We used a
traditional circular Lyot stop with its diameter 80~\% of the pupil.

For taking images, we adjusted the telescope pointing
finely so that the star was
placed at the center of the occulting mask.  Thirty exposures of 0.33~sec
each were coadded into one frame.
Both the telescope and the occulting mask
were dithered by $\sim1"$ every 40 minutes of integration time.
The star was again placed at the center of the
occulting mask, then additional frames were taken. 
The total integration times were
5.9 hour and 1.2 hour for $\epsilon$ Eri and Vega, respectively.

Given the limitations in observation time, 
we did not observe a reference star to determine 
the point spread function (PSF).
As a photometric standard star, 
FS 4 was observed between the observations of Vega and $\epsilon$ Eri.
Dark frames and dome flats with incandescent lamps were taken at the end of the
night.

We observed both objects again on 2004 November 17 with
the same configuration. Integration times for $\epsilon$ Eri and Vega
were 3.5 hour and 12 min, respectively.

The Image Reduction and Analysis Facility (IRAF) was used for data
reduction.  A dark frame was subtracted from each object frame, then
each object frame was divided by the dome-flat.  Hot and bad pixels
were removed from the frame.  

We remove the halo of the central star in each image,
by subtracting the rotated image of the object itself.
The peak position of the PSF moved
slightly on the detector during the observations. This was caused by
the difference in the atmospheric distortion between the infrared
wavelength at which the images were taken and the optical wavelength
at which the wavefront was sensed.  We measured the peak positions
with the RADPROFILE task in IRAF and shifted the images
to adjust the peak position to the center of the image.
Then, each object frame was rotated by 180 degree.
The peak position of the star in the rotated image was slightly
adjusted
so that its wing intensity level is the same as that of the original
image in the region between 1\farcs3 and 2\farcs1 away from the peak.  
In this procedure, some frames were eliminated in which
the AO compensation was poor.
The halo of the star was suppressed in each frame 
after the rotated image was subtracted.
Finally, all frames were combined into one image.

To detect companion candidates, we used the S-Extractor program with a 3$\sigma$
detection threshold above the background.
The extension of the background region strongly affects the source
detection.
We set 32 pixels and 64 pixels as the background mesh sizes,
and an object was assigned to be detected if it was detected with both
background sizes.
We did not count the sources if the ellipticity of PSF calculated by
the program was larger than 0.3 or if the
semi-major axis of its PSF was larger than 3 pixels (0\farcs064).
We also rejected the sources with semi-major axis smaller than 
1 pixel (0\farcs021).

\section{RESULTS AND DISCUSSION}

\subsection{$\epsilon$ Eri}

\begin{figure}
\plotone{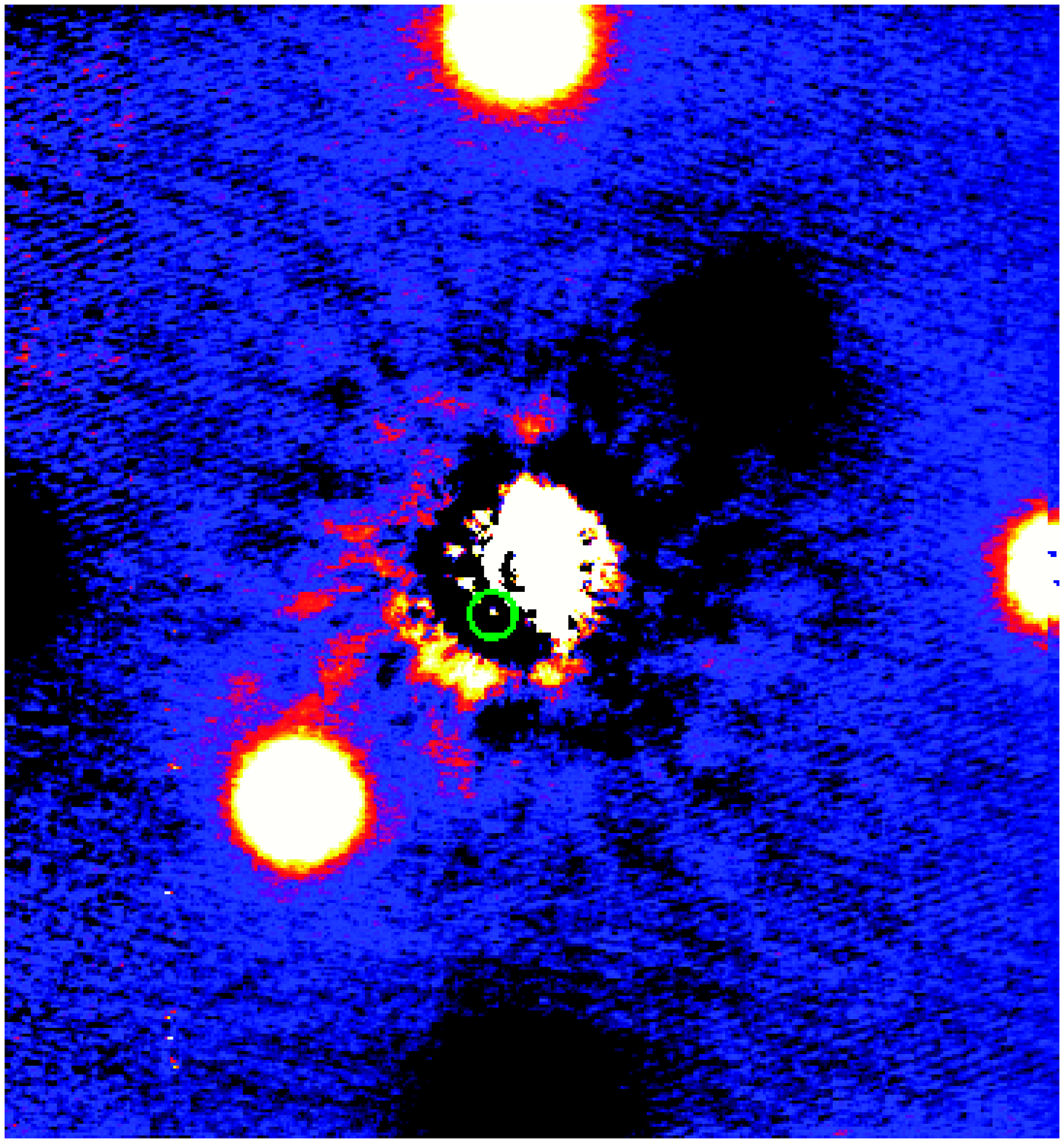}
\caption{An $H$-band coronagraphic image of $\epsilon$ Eri taken
during the 2003 observations presented here.
North is up and east is to the left.
The field of view is 19\farcs6 $\times$ 20\farcs9.
The central star is located at the center of the image but is blocked by 
the mask.
A residual halo from the PSF subtraction remains around the central star.
The other 3 bright emissions are ghosts.
The point source candidate is indicated by a green circle.
\label{eerires}}
\end{figure}

An $H$-band coronagraphic image of $\epsilon$ Eri is presented in
Figure~\ref{eerires}. $\epsilon$ Eri, occulted by the mask, is located at the
center, where the bright speckles display the residual halo of the PSF
subtraction.  
Bright emissions located at the top and the right edge of the image
are ghosts caused by a beam splitter and a compensator of the instrument.
Another ghost at bottom-left to the central star is caused
by the $H$-band filter.

We do not detect diffuse structures around the central star,
whereas a part of the debris disk (Greaves et al. 1998)
is located at the periphery of the field of view.
Proffitt et al. (2004) estimated the surface flux of the disk
to be $\sim 10^{-16}$ erg cm$^{-2}$ s$^{-1}$ arcsec$^{-2}$
at the peak (55 AU)
through optical to near-infrared region.
This was consistent with the upper limit
of the optical HST observations (Proffitt et al. 2004).
Detection limit of our observation is 15.2 mag arcsec$^{-2}$
in the region between 4\arcsec (13 AU) and 10\arcsec (33 AU)
from the star, 4 orders of
magnitude above the predicted flux from the dust.

One faint source is detected in close vicinity of 
the central star in our 2003 image.
Its separation and position angle (PA) relative to the central star are
0\farcs91 and 144\fdg0.
The source has the $H$-band magnitude of $\sim$ 17.3 mag.
We consider this source an artifact,
since the residual halo of PSF subtraction is still dominant
at 0\farcs9 from the central star.
We think that the source is made by an azimuthally inhomogeneous profile
of the PSF of the central star.

Otherwise, it may be an extra-solar planet.
Base on the evolutionary track of low-mass objects (Baraffe et al. 2003), 
the source is estimated to have $6\sim8 M_{\rm J}$, 
if it is associated with $\epsilon$ Eri.
The separation of the source is not inconsistent with
the planet suggested by Hatzes et al. (2000).
As the planet was located near apoastron at the
epoch of the 2003 observation,
its separation from the central star is 4.86 AU (1\farcs5) at most.
Since the planet moved away from its apoastron at the 2004 observation,
it is located too close to the central star to be detected.

The source may be a background star, though a star count model of
the Galaxy (Jones et al. 1987) shows that the expected number
of background stars is only 0.01 within a 5" radius of $\epsilon$ Eri.
Being located close to the Sun, $\epsilon$ Eri has very
large proper motion (0\farcs977/yr with the PA of 271\fdg05).
If the source observed in 2003 is a background star, 
it should be located at 1\farcs71 from the central
star at the 2004 observations.
But no object was identified there.
Bright residual halo of PSF subtraction may prevent us from detecting the source.

\begin{figure}
\plotone{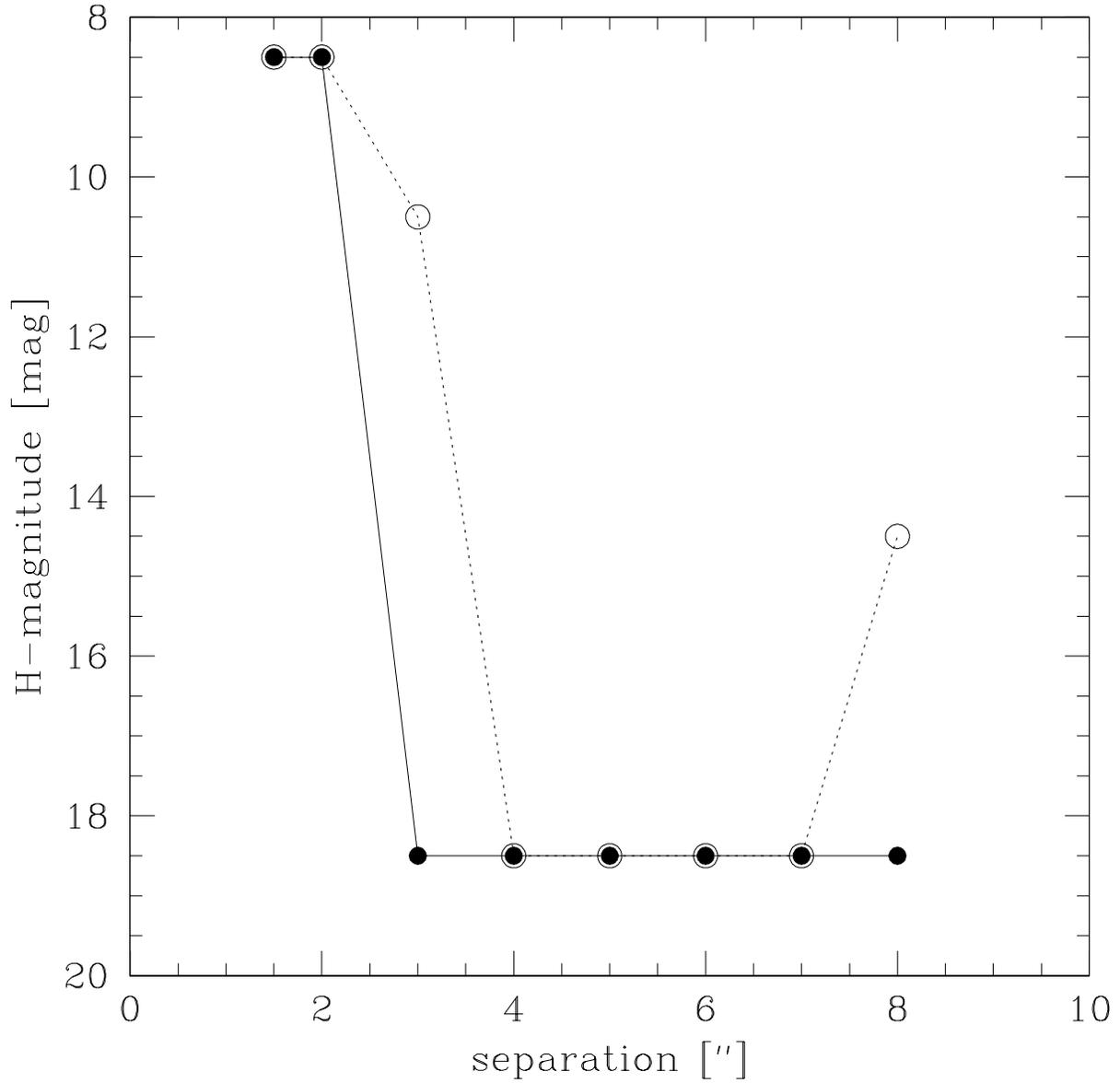}
\caption{
Limiting magnitude of the observation of $\epsilon$ Eri
(filled circles for the 2003 observations, open circles for
the 2004 observations).
The limiting magnitudes are estimated by adding pseudo-stars.
This observation is so far the deepest search for extra-solar planets
at the region between 3\arcsec and 7\arcsec from
the central star.
\label{addstareeri}}
\end{figure}

We estimated the detection limit for a point source by adding pseudo-PSFs
to the raw data.
We made gaussian PSFs with 8.5 mag to 20.5 mag with 2 mag interval and 19.5 mag.
Their FWHMs are 0\farcs075 (3.5 pixel).
Then we placed them at 0\farcs5,
1\farcs0, 1\farcs5, and between 2\farcs0 and 10\farcs0 with 1\farcs0 interval
from the central star.
At each separation, the pseudo-PSFs are located at PA$=-90$\arcdeg, 
$-45$\arcdeg, 0\arcdeg, and $+45$\arcdeg.
When three or four PSFs
at the same separation are identified by the S-Extractor program,
it is assigned to be detected.
The limiting magnitude is shown in Figure $\ref{addstareeri}$,
as a function of the separation from the central star.
At the region between 3$\arcsec$ and 7$\arcsec$ from the central star,
the limiting magnitude is as deep as 18.5 mag at the $H$-band,
corresponding to a $4\sim6 M_{\rm J}$ planet at the same age of $\epsilon$ Eri.
At the region beyond 7$\arcsec$ from the central star, ghosts
near the edges
prohibit us from detecting faint sources.
At the region within 2$\arcsec$ from the central star, 
the detection sensitivity is
severely restricted by the residual halo of the central star.

So far, our observation is the deepest search for extra-solar planets
in the region between 3$\arcsec$ and 7$\arcsec$ from $\epsilon$ Eri.
Several attempts have been made for direct detection of extra-solar
planets around $\epsilon$ Eri.
Macintosh et al. (2003) detected 10 faint objects at 17\arcsec $\sim$ 
45\arcsec away from $\epsilon$ Eri by $K$-band direct imaging observations.
All are beyond the CIAO field of view.
Their following proper
motion measurements indicated that all the objects are background objects.
While the limiting magnitude is about 21.5 mag (corresponding to 
5 $M_{\rm J}$) beyond $15\arcsec$ away from the star,
the sensitivity is poor within $10\arcsec$ from the star.
One reason for such shallow limit
is that they carried out direct imaging observations
without any optics suppressing the brightness of the central star,
such as an occulting mask.

Proffitt et al. (2004) found
59 faint objects in the region between 12\farcs5 and $58\arcsec$ from
$\epsilon$ Eri. 
Most of them are elongated, suggestive of background galaxies.
They did not detect any object within our field of view.
Though the detection limit of their observation is as deep as
26 mag in optical wavelengths,
extra-solar planets are estimated to be orders of magnitude fainter
in optical wavelengths than their detection limit.

Structure of the debris disk might be influenced by a planet. 
Quillen \& Thorndike (2002) predicts a giant planet with a
semi-major axis of 40 AU. 
Such a planet may be located beyond the CIAO field of view.
Otherwise, a planet might be less massive.
With an evolutionary track of Baraffe et al. (2003),
a 1 $M_{\rm J}$ planet is expected to be
as faint as 26 mag to 30 mag at the $H$-band.
The negative result of our observation
is therefore not inconsistent with their prediction.

\subsection{Vega}

\begin{figure}
\plotone{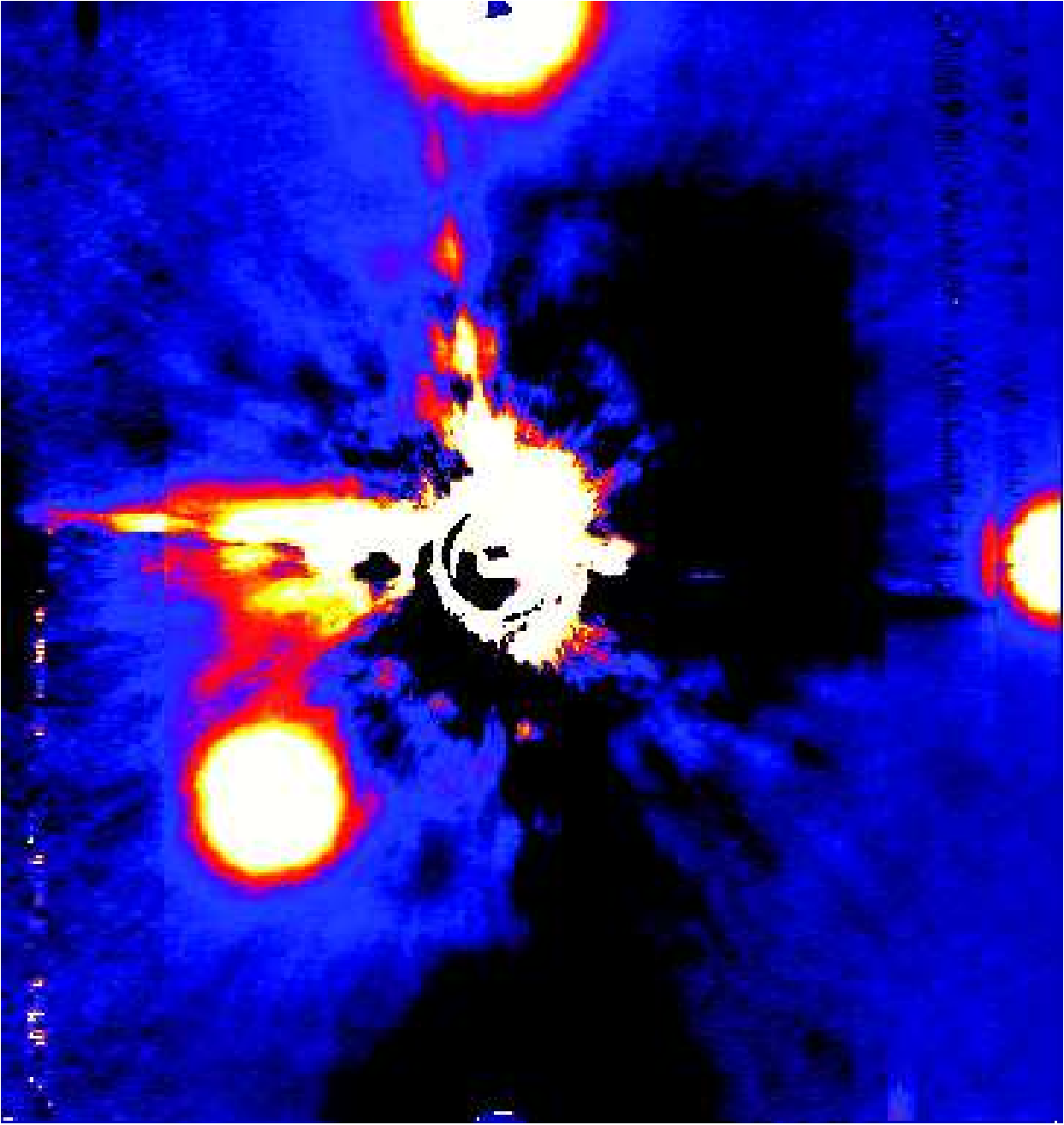}
\caption{An $H$-band coronagraphic image of Vega taken during the 2003
observations.
The field of view is 18\farcs9 $\times$ 20\farcs2.
The star is located at the center of the image.
The emission structures are residuals of the PSF subtraction, 
diffraction patterns of the spider, and ghosts.
No stellar object was found around Vega.
\label{vegares}}
\end{figure}

An $H$-band coronagraphic image of Vega is presented in
Figure~\ref{vegares}. Vega, occulted by the mask, is located at the
center. 
We do not detect any diffuse circumstellar structure,
though two dust peaks identified by the sub-mm observations (Wilner et al. 2002)
are located within the field of view of CIAO.
Detection limit of our observation is 15 mag arcsec$^{-2}$
around the dust peaks.

\begin{figure}
\epsscale{1.0}
\plotone{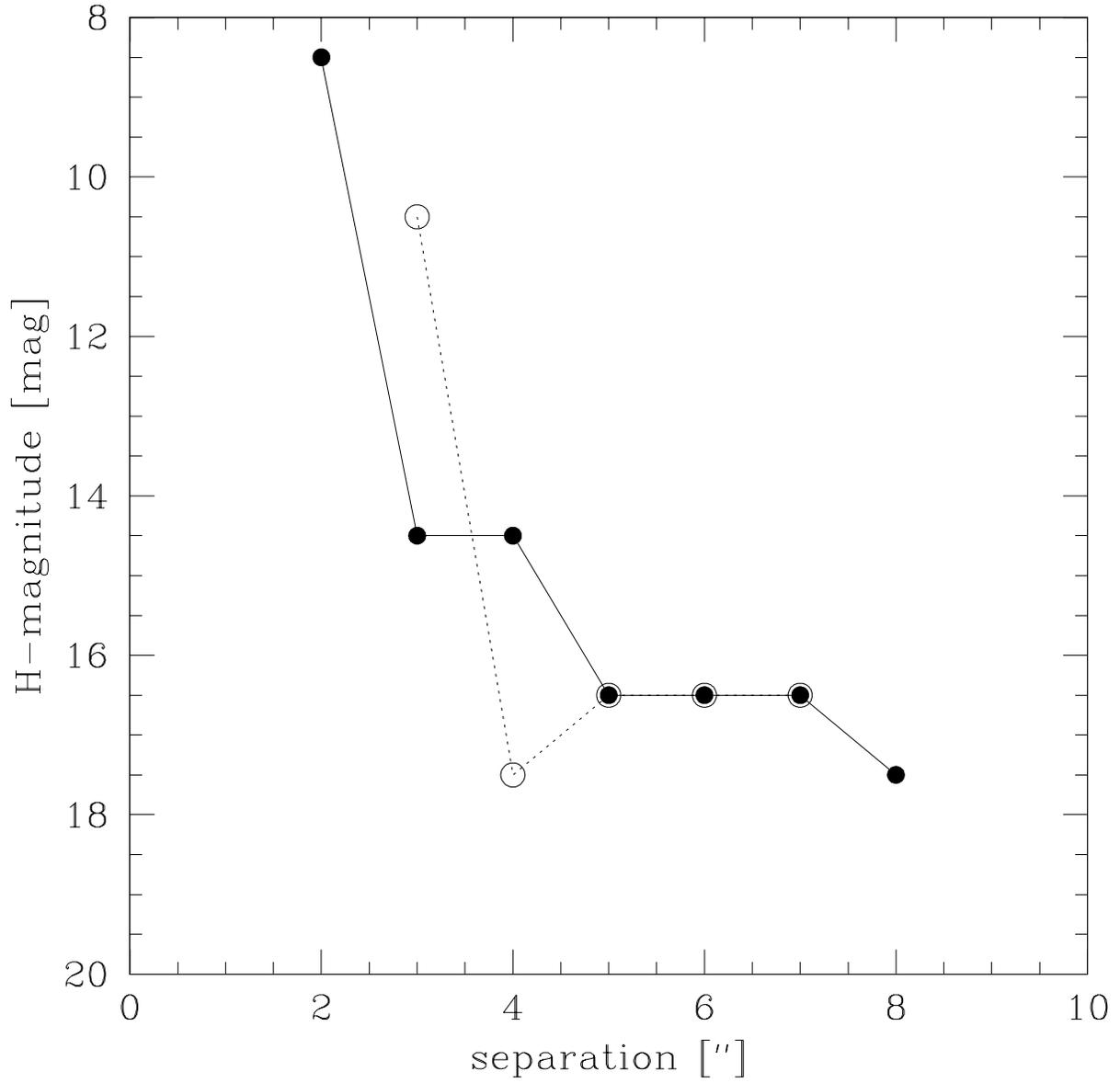}
\caption{
Limiting magnitude of the observation of Vega.
The symbols are the same as in Figure \ref{addstareeri}.
The limiting magnitudes are estimated by adding artificial PSFs.
\label{addstarvega}}
\end{figure}

We do not detect any point source
around the central star in both epochs.
We estimate the detection limit for a point source by the same procedure 
we used for the $\epsilon$ Eri data (Figure $\ref{addstarvega}$).
The limiting magnitude ($\sim 17$ mag between 5\arcsec and 8\arcsec)
is shallower than that for $\epsilon$ Eri.
This is because integration time is shorter than that for
$\epsilon$ Eri and Vega is much brighter
than $\epsilon$ Eri.
On the other hand, the detection limit in terms of mass is $5\sim10 M_{\rm J}$,
similar to that of $\epsilon$ Eri, because Vega is younger than 
$\epsilon$ Eri.

A planet could induce inhomogeneity in the dust distribution.
A planet, with Neptunian mass or several Jupiter mass, 
is predicted at 50 AU (6\farcs4) or 65 AU (8\farcs4) away
from the central star (Ozernoy et al. 2000; Wilner et al. 2002;
Wyatt et al. 2003).
The negative result of our observation constrains the mass of the planet that 
they predict to less than $5\sim10 M_{\rm J}$.

An extra-solar planet around Vega has also been investigated.
Macintosh et al. (2003) searched extra-solar planets
also around Vega by direct imaging observations.
Their $K$-band limiting magnitude was $\sim$ 20.5 mag beyond 20$\arcsec$ 
from the central star,
while only $\sim$ 17 mag ($6\sim12 M_{\rm J}$)
at 7\arcsec from the central star.
7 objects were found $>$ 20\arcsec away from the central star.
Based on the proper motion measurements, they are thought to be
background stars.
Metchev et al. (2003) also investigated extra-solar planets around Vega.
Their $H$-band limiting magnitude was about 19 mag ($2\sim6 M_{\rm J}$)
and 14 mag ($10\sim20 M_{\rm J}$) at $20\arcsec$ and 
$7\arcsec$ from the central star, respectively.
They detected 8 background stars.

Marois et al. (2006) observed Vega at 1.6 $\mu$m using the recently developed
method for high contrast imaging.
While their limiting magnitude reaches as deep as 20 mag at 8\arcsec
offset from the central star,
they did not detect any faint object around Vega.
Hinz et al. (2006) carried out $M$-band direct imaging observations of 
Vega. Using AO, they obtained diffraction-limited images
with a detection limit of 7 $M_{\rm J}$ at 2\farcs5 from the central star.

With a limiting magnitude that corresponds to $5\sim10 M_{\rm J}$ in
the region between 5\arcsec and 8\arcsec from the central star,
our observations are so far one of
the deepest search for an extra-solar planet around
Vega, 

\section{CONCLUSIONS}

We have carried out near-infrared coronagraphic observations of
$\epsilon$ Eri and Vega.
The observations are one of the deepest near-infrared
searches for extra-solar 
planets in close vicinity of both central stars so far.
\begin{enumerate}
\item We did not detect any trustworthy point source around
$\epsilon$ Eri. 
The upper limit on the mass of a planet is $4\sim6 M_{\rm J}$
in the region between 3\arcsec (10 AU) and 7\arcsec (23 AU) from 
the star.
Location of one point source candidate
is not inconsistent with the planet suggested by the Doppler shift 
measurements.
Another epoch observation will make it clear whether the candidate is a 
real object and whether it orbits $\epsilon$ Eri.
\item We did not detect any point source around Vega.
Negative result of our observations puts an upper limit on the mass of 
a planet to $5\sim10 M_{\rm J}$ in
the region between 5\arcsec (40 AU) and 8\arcsec (60 AU) from the star.
\end{enumerate}

\acknowledgments

We thank Miki Ishii and Sumiko Harasawa for the support
for making observations and Ingrid Mann for careful reading of the manuscript.
This work was supported by ``The 21st
Century COE Program: The Origin and Evolution of Planetary Systems" of
the Ministry of Education, Culture, Sports, Science and Technology
(MEXT).  Y. I. is supported by the Sumitomo Foundation, the Ito
Science Foundation, and a Grant-in-Aid for Scientific Research
No. 16740256 of the MEXT.

\clearpage

%

\end{document}